\documentclass[12pt]{article}%
\usepackage{amsmath}
\usepackage{amsfonts}
\usepackage{amssymb}
\usepackage{graphicx}%
\providecommand{\U}[1]{\protect\rule{.1in}{.1in}}
\providecommand{\U}[1]{\protect\rule{.1in}{.1in}}
\providecommand{\U}[1]{\protect\rule{.1in}{.1in}}
\providecommand{\U}[1]{\protect\rule{.1in}{.1in}}
\providecommand{\U}[1]{\protect\rule{.1in}{.1in}}
\providecommand{\U}[1]{\protect\rule{.1in}{.1in}}
\providecommand{\U}[1]{\protect\rule{.1in}{.1in}}
\providecommand{\U}[1]{\protect\rule{.1in}{.1in}}
\providecommand{\U}[1]{\protect\rule{.1in}{.1in}} \textwidth17.1cm \oddsidemargin -0.5cm

\newcommand{\be}{\begin{equation}}
\newcommand{\ee}{\end{equation}}
\newcommand{\ben}{\begin{equation*}}
\newcommand{\een}{\end{equation*}}
\newcommand{\ar}{\begin{array}}
\newcommand{\arn}{\end{array}}

\def\pnot{\mbox{${\not{\hbox{\kern-3.0pt$p$}}}$}}
\def\qnot{\mbox{${\not{\hbox{\kern-2.0pt$q$}}}$}}
\def\enot{\mbox{${\not{\hbox{\kern-2.0pt$e$}}}$}}
\def\knot{\mbox{${\not{\hbox{\kern-2.0pt$k$}}}$}}

\def\fun#1#2{\lower3.6pt\vbox{\baselineskip0pt\lineskip.9pt\ialign
{$\mathsurround=0pt#1\hfil##\hfil$\crcr#2\crcr\sim\crcr}}}

\begin{document}
\begin{titlepage}
\begin{center}
{\bf Matching of the low-x evolution kernels$^{~\ast}$}
\end{center}
\vskip 0.5cm \centerline{V.S.~Fadin$^{a\,\dag}$,
R.~Fiore$^{b\,\ddag}$, A.V.~Grabovsky$^{a\,\dag\dag}$} \vskip .6cm
\centerline{\sl $^{a}$ Budker Institute of Nuclear Physics, 630090
Novosibirsk, Russia} \centerline{\sl Novosibirsk State University,
630090 Novosibirsk, Russia} \centerline{\sl $^{b}$ Dipartimento di
Fisica, Universit\`a della Calabria,} \centerline{\sl Istituto
Nazionale di Fisica Nucleare, Gruppo collegato di Cosenza,}
\centerline{\sl Arcavacata di Rende, I-87036 Cosenza, Italy}
\vskip 2cm
\begin{abstract}
We demonstrate that the  ambiguity of the low-x evolution kernels in the next-to-leading order (NLO) permits one to match  the M\"{o}bius form of  BFKL kernel and  the kernel of the colour dipole model and  to  construct the M\"{o}bius invariant NLO BFKL kernel in $N=4$ supersymmetric Yang-Mills theory.
\end{abstract}
\vfill \hrule \vskip.3cm \noindent $^{\ast}${\it Work supported in
part by the RFBR grant 07-02-00953, in part by the RFBR--MSTI
grant 06-02-72041, in part by the INTAS
grant 05-1000008-8328 and in part by Ministero Italiano
dell'Istruzione, dell'Universit\`a e della Ricerca.} \vfill $
\begin{array}{ll} ^{\dag}\mbox{{\it e-mail address:}} &
\mbox{FADIN@INP.NSK.SU}\\
^{\ddag}\mbox{{\it e-mail address:}} &
\mbox{FIORE@CS.INFN.IT}\\
^{\dag\dag}\mbox{{\it e-mail address:}} &
\mbox{A.V.GRABOVSKY@INP.NSK.SU}\\
\end{array}
$
\end{titlepage}

\vfill \eject

\section{Introduction}

This paper is devoted to the investigation of conformal properties of the
Balitsky-Fadin-Kuraev-Lipatov (BFKL) equation \cite{BFKL} and the relationship
between the BFKL approach and the colour dipole model \cite{dipole}.

The BFKL approach gives the most common basis for the theoretical description
of semihard processes in QCD and has a wide field of applications covering
scattering processes with arbitrary momentum and colour exchanges. Now the
next-to-leading (NLO) corrections to the kernel of the BFKL equation are known
both for the forward scattering \cite{Fadin:1998py}, \cite{Ciafaloni:1998gs}
and for any momentum and colour transfer \cite{Fadin:1998jv}-\cite{FF05}.

The BFKL approach is based on the gluon reggeization. Originally it was
formulated in the momentum representation, and the BFKL kernel was calculated
in the space of transverse momenta $\vec q_{1}, \vec q_{2}$ of two interacting
reggeized gluons. Later it was recognized that for the case of scattering of
colourless objects the BFKL equation possesses remarkable properties, which
become mostly apparent in the space of conjugate coordinates $\vec r_{1}, \vec
r_{2}$. It was shown \cite{Lipatov:1985uk} that in this case the BFKL equation
can be written in the special representation (in the space of states
$|\Psi\rangle$ with the ``dipole" property $\langle\vec r, \vec r|\Psi
\rangle=0$), where the equation is invariant under the conformal (M\"{o}bius)
transformations of the transverse coordinates. Following
Ref.~\cite{Bartels:2004ef} we will call this representation M\"{o}bius
representation. For brevity, we will also call the BFKL kernel in this
representation M\"{o}bius kernel, and its form in the coordinate space
M\"{o}bius (or dipole) form. The M\"{o}bius form of the LO BFKL kernel is
explicitly conformal invariant \cite{Fadin:2006ha}. Moreover, it coincides
with the kernel of the evolution equation in the colour dipole model.

The colour dipole model is formulated in the coordinate space. Unlike the BFKL
approach, it is applicable only to scattering of colorless particles. Its
attractive feature is a clear physical interpretation. The model is applied
not only at low parton density, but also in the high density regime, where
parton fusion is essential \cite{Gribov:1984tu}, and evolution equations
become nonlinear. In general, in this regime there is an infinite hierarchy of
coupled equations \cite{Balitsky:1995ub}--\cite{CGC}. In the simplest case,
when the target is a large nucleus, it is reduced to the Balitsky-Kovchegov
(BK) equation \cite{Kovchegov:1999yj}. It was shown \cite{Bartels:2004ef} that
the BK equation appears as a special case of the nonlinear evolution equation
which sums the fan diagrams for the BFKL Green's functions in the M\"{o}bius
representation. Therefore in the LO there is a full agreement between the BFKL
approach and colour dipole model.

Recently the NLO corrections to the BK kernel have been calculated
\cite{{Kovchegov:2006vj}}--\cite{Balitsky:2008zz} and investigation of
inter-relation of the BFKL approach and the colour dipole model in the NLO
became possible. A clear understanding of this inter-relation is important for
the further development of theoretical description of small-x processes. Not
less significant is the understanding of conformal properties of the NLO BFKL
kernel in the M\"{o}bius representation because the conformal invariance is
extremely important for integrability of the BFKL equation. Evidently, in QCD
the conformal invariance is violated by the running coupling, i.e. by the
terms proportional to $\beta$-- function. But one could expect that the
M\"{o}bius form is quasi-conformal, i.e. the conformal invariance is violated
only by such terms, and that it remains unbroken in $N=4$ SUSY Yang-Mills One
could expect also coincidence of the M\"{o}bius form of the BFKL kernel and
the kernel of the colour dipole model.

However, the situation is not so simple, because the NLO kernels are not
unambiguously defined. Their ambiguity is analogous to the ambiguity of the
NLO anomalous dimensions. It is caused by the possibility to redistribute
radiative corrections between the kernels and the impact factors.

We prove that this ambiguity permits one to match the M\"{o}bius form of the
BFKL kernel and the kernel of the colour dipole model and to construct the
M\"{o}bius invariant NLO BFKL kernel in $N=4$ SUSY.

The NLO corrections consist of quark and gluon parts. The quark part of the BK
kernel was found Refs.~\cite{Kovchegov:1999yj, Balitsky:2006wa}; the
corresponding part of the M\"{o}bius form of the BFKL kernel was calculated in
Refs.~\cite{Fadin:2006ha,Fadin:2007ee}. Taking into account of the ambiguity
mentioned above, it was shown there that up to the difference in the
renormalization scales the quark parts agree with each other. Moreover, the
\textquotedblleft abelian" piece of the quark part is conformal invariant
\cite{Fadin:2007ee}. This is especially interesting for the QED Pomeron
\cite{Gribov:1970ik, Cheng:1970xm} because this piece is proportional to the
total QED kernel.

The M\"{o}bius form for the gluon part was obtained in
Ref.~\cite{Fadin:2007de}. As well as the quark part, it turned out strikingly
simple compared with the gluon part of the BFKL kernel in transverse momentum
space. However, it was found that the conformal invariance of this form is
broken not only by the terms proportional to $\beta$-- function. In principle,
this result did not mean that the conformal invariance was broken in $N=4$
SUSY Yang-Mills theory because the field structure of this theory differs from
QCD. The extension of the BFKL framework to the supersymmetric theories was
started in Ref.~\cite{Kotikov:2000pm}, where the forward kernel was found for
the SUSY N=4 in the space of the Born eigenfunctions and in the momentum space
with the dimension $D=4+2\epsilon$. This analysis has been expanded Ref.~the
\cite{Fadin:2007xy}, where the M\"{o}bius form of the nonforward BFKL kernel
was obtained for the supersymmetric theories with arbitrary N. It turned out
that this form violates conformal invariance at any N. However, because of the
ambiguity discussed above a possibility of existence of a conformal invariant
kernel at $N=4$ was not excluded. Also Regge limits of 4-point correlators were
studied in N=4 SUSY directly in the coordinate space via conformal invariance
and the connection of the analysis to the AdS/CFT duality was discussed
in Ref.~\cite{cornalba}.

The gluon part of the BK kernel calculated in Ref.~\cite{Balitsky:2008zz}
agreed neither with the M\"{o}bius form of Ref.~\cite{Fadin:2007de}, nor with
the eigenvalues of the forward BFKL kernel \cite{Fadin:1998py, Kotikov:2000pm}%
. It was found afterwards that Ref.~\cite{Balitsky:2008zz} contained an error,
which was corrected in Ref.~\cite{Balitsky:2009xg}. The discrepancy of the
results of Refs.~\cite{Balitsky:2008zz} and \cite{Fadin:1998py,
Kotikov:2000pm, Fadin:2007de} was analyzed in detail in
Ref.~\cite{Fadin:2009za}. A special attention was paid to the case of the
forward scattering. It was shown for this case that with account of the
ambiguity of the NLO kernels, of the correction performed in
Ref.~\cite{Balitsky:2009xg} and of the difference in the renormalization
scales, the discrepancy disappeared. Besides this, the functional identity of
the forward BFKL kernel in the momentum and M\"{o}bius representations in the
leading order (LO) was exhibited and its NLO validity in $N=4$ supersymmetric
Yang-Mills theory was proved.

Here we demonstrate that with account of the correction of
Ref.~\cite{Balitsky:2009xg} and the difference in the renormalization scales
(it was discussed in detail in Ref.~\cite{Fadin:2009za} and we will not
mention it further) the ambiguity of the NLO kernels permits us to match the
gluon parts for the non-forward case also. Moreover, this ambiguity allows us
to present the kernel in the form where the conformal invariance is violated
only by renormalization. It is especially interesting for the N=4 SUSY Yang-Mills.

Our paper is organized as follows. In the next section, our notation is
introduced, the ambiguity of the NLO kernels is briefly discussed and a short
overview of the discrepancy between the kernels is given. Section 3 presents
the transformation of the BFKL kernel (allowed by the NLO ambiguity) which
removes this discrepancy. In section 4 the transformation to the
quasi-conformal shape is presented. Section 5 gives the quasi-conformal
(conformal at N=4) kernels in QCD and SUSY Yang-Mills theories. Section 6
presents our conclusions.

\section{General overview}

\label{sec:notation} Our notation is the same as in Refs.~
\cite{Fadin:2006ha,Fadin:2007de}. Thus we denote Reggeon transverse momenta
(conjugate coordinates) in initial and final $t$-channel states as $\vec
{q}_{i}^{\;\prime}\; (\vec{r}_{i}^{\;\prime})$ and $\vec{q}_{i}$ ($\vec{r}%
_{i}),$ $i=1,2$. At space-time dimension $D=4+2\epsilon$ the state
normalization is
\begin{equation}
\langle\vec{q}|\vec{q}^{\;\prime}\rangle=\delta(\vec{q}-\vec{q}^{\;\prime
})\;,\;\;\;\;\;\langle\vec{r}|\vec{r}^{\;\prime}\rangle=\delta(\vec{r}-\vec
{r}^{\;\prime})\;,\quad\quad\langle\vec{r}|\vec{q}\rangle=\frac{e^{i\vec
{q}\,\vec{r}}}{(2\pi)^{(1+\epsilon)}}. \; \label{normalization}%
\end{equation}
We will also use $\vec{p}_{ij^{\prime}}=\vec{p}_{i}-\vec{p}_{j}^{\;\prime}$
for brevity.

The $s$-channel discontinuities of scattering amplitudes for the processes
$A+B\rightarrow A^{\prime}+B^{\prime}$ have the form
\begin{equation}
-4i(2\pi)^{D-2}\delta(\vec{q}_{A}-\vec{q}_{B})\mbox{disc}_{s}\mathcal{A}%
_{AB}^{A^{\prime}B^{\prime}}=\langle A^{\prime}\bar{A}|\left(  \frac{s}{s_{0}%
}\right)  ^{\hat{\mathcal{K}}}\frac{1}{\hat{\vec{q}}_{1}^{\;2}\hat{\vec{q}%
}_{2}^{\;2}}|\bar{B}^{\prime}B\rangle\;. \label{discontinuity representation}%
\end{equation}
In this expression $s_{0}$ is an appropriate energy scale, $\;\;q_{A}%
=p_{A^{\prime}A},\;\;q_{B}=p_{BB^{\prime}}$, and $\hat{\mathcal{K}} $\ is the
BFKL kernel,
\begin{equation}
\langle\vec{q}_{1},\vec{q}_{2}|\mathcal{\hat{K}\,}|\vec{q}_{1}^{\,\,\prime
},\vec{q}_{2}^{\,\,\prime}\rangle=\delta(\vec{q}_{11^{\prime}}+\vec
{q}_{22^{\prime}})\frac{\mathcal{K}_{r}\left(  q_{1},q_{1}^{\prime},q\right)
}{\vec{q}_{1}^{\,\,2}\vec{q}_{2}^{\,\,2}}+\delta(\vec{q}_{22^{\prime}}%
)\delta\left(  \vec{q}_{11^{\prime}}\right)  \left(  \omega\left(  \vec{q}%
_{1}^{\,\,2}\right)  +\omega(\vec{q}_{2}^{\,\,2})\right)  ,
\label{K_momentum_repr_definition}%
\end{equation}
where $\omega(t)$ is the gluon Regge trajectory and $\mathcal{K}_{r}\left(
q_{1},q_{1}^{\prime},q\right) $ represents real particle production in Reggeon
collisions. The impact factors are introduced through
\begin{equation}
\langle\vec{q}_{1},\vec{q}_{2}|\bar{B}^{\prime}B\rangle=4p_{B}^{-}\delta
(\vec{q}_{B}-\vec{q}_{1}-\vec{q}_{2}){\Phi}_{B^{\prime}B}(\vec{q}_{1},\vec
{q}_{2})\;, \label{impact BB}%
\end{equation}%
\begin{equation}
\langle A^{\prime}\bar{A}|\vec{q}_{1},\vec{q}_{2}\rangle=4p_{A}^{+}\delta
(\vec{q}_{A}-\vec{q}_{1}-\vec{q}_{2}){\Phi}_{A^{\prime}A}(\vec{q}_{1},\vec
{q}_{2})\;, \label{impact AA}%
\end{equation}
where $p^{\pm}=(p_{0}\pm p_{z})/\sqrt{2}$. The kernel $\mathcal{K}_{r}(\vec
{q}_{1},\vec{q}_{1}^{\;\prime};\vec{q})$ and the impact factors $\Phi$ are
expressed through the Reggeon vertices according to Ref.~\cite{FF98}. So, the
real part of the Born kernel reads
\begin{equation}
\frac{\mathcal{K}_{r}^{B}(\vec{q}_{1},\vec{q}_{1}^{\;\prime};\vec{q})}{\vec
{q}_{1}^{\,\,2}\vec{q}_{2}^{\,\,2}}=\frac{\alpha_{s}(\mu^{2})N_{c}}{\pi^{2}%
}\left(  \frac{1}{\vec{q}_{11^{\prime}}^{\,\,2}}+\frac{\left(  \vec{q}%
_{2}\,\vec{q}_{11^{\prime}}\right)  }{\vec{q}_{2}^{\,\,2}\vec{q}_{11^{\prime}%
}^{\,\,2}}-\frac{\left(  \vec{q}_{1}\,\vec{q}_{11^{\prime}}\right)  }{\vec
{q}_{1}^{\,\,2}\vec{q}_{11^{\prime}}^{\,\,2}}-\frac{\left(  \vec{q}_{1}%
\,\vec{q}_{2}\right)  }{\vec{q}_{1}^{\,\,2}\vec{q}_{2}^{\,\,2}}\right)  ,
\end{equation}
while the one-loop trajectory has the form%
\begin{equation}
\omega(\vec{q}^{\;2})=-\frac{\alpha_{s}(\mu^{2})N_{c}}{(2\pi)^{2+2\epsilon}%
}\int d^{2+2\epsilon}k\left(  \frac{2}{\vec{k}^{\,\,2}}-\frac{2\vec{k}(\vec
{k}-\vec{q})}{\vec{k}^{\,\,2}(\vec{k}-\vec{q})^{2}}\right)  .
\end{equation}
At once we can see that Eq.~(\ref{discontinuity representation})\ does not
give the unique definition of the kernel. Indeed, the discontinuity
$\mbox{disc}_{s}\mathcal{A}_{AB~}^{A^{\prime}B^{\prime}}$ in
Eq.~(\ref{discontinuity representation}) remains intact if one changes both
the kernel and the impact factors via an arbitrary nonsingular operator:
$\hat{\mathcal{O}}$
\begin{equation}
~ \hat{\mathcal{K}}\rightarrow\hat{\mathcal{O}}^{-1}\hat{\mathcal{K}}%
\hat{\mathcal{O}}~,\;\;\langle A^{\prime}\bar{A}|\rightarrow\langle A^{\prime
}\bar{A}|\hat{\mathcal{O}}~,\;\;\frac{1}{\hat{\vec{q}}_{1}^{\;2}\hat{\vec{q}%
}_{2}^{\;2}}|\bar{B}^{\prime}B\rangle\rightarrow{\hat{\mathcal{O}}^{-1}}%
\frac{1}{\hat{\vec{q}}_{1}^{\;2}\hat{\vec{q}}_{2}^{\;2}}|\bar{B}^{\prime
}B\rangle. \label{kernel transformation}%
\end{equation}
Actually, the kernel $\hat{\mathcal{K}}$ (\ref{K_momentum_repr_definition}) is
obtained from the symmetric kernel, usually used in the momentum space, just
by such transformation. Only owing to this transformation the M\"{o}bius form
of $\hat{\mathcal{K}}$ is conformal invariant and coincides with the dipole
kernel in the leading order. But even if the kernel is fixed in the LO,
transformations with $\hat{{O}}=1-\hat{O}$, where $\hat{O}\sim\alpha_{s}$, are
still possible. Within the NLO accuracy these transformations give
\begin{equation}
\hat{\mathcal{K}}\rightarrow\hat{\mathcal{K}}-[\hat{\mathcal{K}}^{B},\hat{O}].
\label{kernel_transforms}%
\end{equation}
Such transformations can be used for simplify the form of the kernel, in
particular of its M\"{o}bius form. Indeed, it was shown \cite{Fadin:2006ha,
Fadin:2007de} that this form is simplified by the transformation
\begin{equation}
\hat{\mathcal{K}}\rightarrow\hat{\emph{K}}= \hat{\mathcal{K}}+\frac{\alpha
_{s}}{8\pi}\beta_{0} [\hat{\mathcal{K}}^{B}, \ln\left( \hat{\vec{q}}_{1}%
^{\;2}\hat{\vec{q}}_{2}^{\;2}\right) ]~ , \label{transform with beta}%
\end{equation}
where $\hat{\mathcal{K}}$ is the kernel defined in Eq.~
(\ref{K_momentum_repr_definition}), $\hat{\mathcal{K}}^{B}=\hat{\emph{K}%
}^{\:B}$ is its LO value and $\beta_{0}$ is the first coefficient of the
beta-function. In the following this transformation is assumed to be done.

In the NLO the M\"{o}bius form can be written \cite{Fadin:2006ha,Fadin:2007de}
as follows:%
\[
\langle\vec{r}_{1},\vec{r}_{2}|\hat{\emph{K}}_{M}|\vec{r}_{1}^{\;\prime}%
,\vec{r}_{2}^{\;\prime}\rangle=\frac{\alpha_{s}(\mu^{2})N_{c}}{2\pi^{2}}\int
d\vec{\rho}\frac{\vec{r}_{12}{}^{2}}{\vec{r}_{1\rho}^{\,\,2}\vec{r}_{2\rho
}^{\,\,2}}\Biggl[\delta(\vec{r}_{11^{\prime}})\delta(\vec{r}_{2^{\prime}\rho
})+\delta(\vec{r}_{1^{\prime}\rho})\delta(\vec{r}_{22^{\prime}})-\delta
(\vec{r}_{11^{\prime}})\delta({r}_{22^{\prime}})~\Biggr]
\]%
\[
+\frac{\alpha_{s}^{2}(\mu^{2})N_{c}^{2}}{4\pi^{3}}\Biggl[\delta(\vec
{r}_{11^{\prime}})\delta(\vec{r}_{22^{\prime}})\int d\vec{\rho}\,g_{0}(\vec
{r}_{1},\vec{r}_{2};\rho)+\delta(\vec{r}_{11^{\prime}})g_{1}(\vec{r}_{1}%
,\vec{r}_{2};\vec{r}_{2}^{\;\prime})+\delta(\vec{r}_{22^{\prime}})g_{1}%
(\vec{r}_{2},\vec{r}_{1};\vec{r}_{1}^{\;\prime})
\]%
\begin{equation}
+\frac{1}{\pi}g_{2}(\vec{r}_{1},\vec{r}_{2};\vec{r}_{1}^{\;\prime},\vec{r}%
_{2}^{\;\prime})\Biggr]. \label{K_in_corrdinate_repr}%
\end{equation}
Here $\vec{r}_{i\rho}=\vec{r}_{i}-\vec{\rho}$, and the whole kernel is
symmetric with respect to the substitution $1\leftrightarrow2, 1^{\prime
}\leftrightarrow2^{\prime}$. The M\"{o}bius kernel (\ref{K_in_corrdinate_repr}%
) is defined with an accuracy to any functions independent of $\vec{r}_{1}$ or
of $\vec{r}_{2}$ such that after their addition the kernel remains zero at
$\vec{r}_{1}=\vec{r}_{2}$ $\cite{Fadin:2006ha,Fadin:2007de}.$ Therefore, one
can add to the kernel only the functions which are antisymmetric with respect
to the $\vec{r}_{1}\leftrightarrow\vec{r}_{2}$ substitution. These functions
do not change the symmetric part of the kernel. But this part alone plays a
role because of the symmetry of the impact factors.

The transformation (\ref{transform with beta}) considerably simplifies the
\textquotedblleft non-abelian" piece of the quark part of the M\"{o}bius form
\cite{Fadin:2006ha}. In particular, it removes its contribution to the
function $g_{2}$. Moreover, just after this transformation the M\"{o}bius form
of the quark contribution to the BFKL kernel \cite{Fadin:2006ha, Fadin:2007ee}
coincides with the quark part of the linearized BK kernel
\cite{Kovchegov:2006vj, Balitsky:2006wa} calculated in the colour dipole model.

However, the transformation (\ref{transform with beta}) does not remove the
disagreement of the gluon contribution. The BFKL framework gives for the gluon
contribution to the functions $g_{i}$ in (\ref{K_in_corrdinate_repr})
\cite{Fadin:2007de}
\begin{equation}
g_{0}(\vec{r}_{1},\vec{r}_{2};\vec{\rho})=2\pi\zeta(3)\delta\left(  \vec{\rho
}\right)  -g(\vec{r}_{1},\vec{r}_{2};\vec{\rho})\ ,\label{g0BFKL}%
\end{equation}%
\[
g_{1}(\vec{r}_{1},\vec{r}_{2};\vec{r}_{2}^{\;\prime})\ =\frac{11}{6}\frac
{\vec{r}_{12}^{\;2}}{\vec{r}_{22^{\prime}}^{\;2}\vec{r}_{12^{\prime}}^{\;2}%
}\ln\left(  \frac{\vec{r}_{12}^{\;2}}{r_{\mu}^{2}}\right)  +\frac{11}%
{6}\left(  \frac{1}{\vec{r}_{22^{\prime}}^{\,\,\,2}}-\frac{1}{\vec
{r}_{12^{\prime}}^{\,\,\,2}}\right)  \ln\left(  \frac{\vec{r}_{22^{\prime}%
}^{\,\,\,2}}{\vec{r}_{12^{\prime}}^{\,\,\,2}}\right)
\]%
\begin{equation}
+\frac{1}{2\vec{r}_{22^{\prime}}^{\;2}}\ln\left(  \frac{\vec{r}_{12^{\prime}%
}^{\;2}}{\vec{r}_{22^{\prime}}^{\;2}}\right)  \ln\left(  \frac{\vec{r}%
_{12}^{\;2}}{\vec{r}_{12^{\prime}}^{\;2}}\right)  -\frac{\vec{r}_{12}^{\;2}%
}{2\,\vec{r}_{22^{\prime}}^{\;2}\vec{r}_{12^{\prime}}^{\;2}}\ln\left(
\frac{\vec{r}_{12}^{\;2}}{\vec{r}_{22^{\prime}}^{\;2}}\right)  \ln\left(
\frac{\vec{r}_{12}^{\;2}}{\vec{r}_{12^{\prime}}^{\;2}}\right)  ,\label{g1BFKL}%
\end{equation}
where
\begin{equation}
\ln r_{\mu}^{2}=2\psi\left(  1\right)  -\ln\frac{\mu^{2}}{4}-\frac{3}%
{11}\left(  \frac{67}{9}-2\zeta(2)\right)  ,\label{rmu}%
\end{equation}
and
\begin{align}
g_{2}(\vec{r}_{1},\vec{r}_{2};\vec{r}_{1}^{\;\prime},\vec{r}_{2}^{\;\prime})
&  =\frac{1}{2\vec{r}_{1^{\prime}2^{\prime}}^{\,\,4}}\left(  \frac{\vec
{r}_{11^{\prime}}^{\;2}\,\vec{r}_{22^{\prime}}^{\;2}-2\vec{r}_{12}^{\;2}%
\,\vec{r}_{1^{\prime}2^{\prime}}^{\;2}}{d}\ln\left(  \frac{\vec{r}%
_{12^{\prime}}^{\;2}\,\vec{r}_{21^{\prime}}^{\;2}}{\vec{r}_{11^{\prime}}%
^{\;2}\vec{r}_{22^{\prime}}^{\;2}}\right)  -1\right)  +\frac{\vec{r}%
\,\,_{12}^{2}\ln\ \left(  \frac{\vec{r}\,\,_{11^{\prime}}^{2}}{\vec
{r}\,\,_{1^{\prime}2^{\prime}}^{2}}\right)  }{2\vec{r}\,\,_{11^{\prime}}%
^{2}\vec{r}\,\,_{12^{\prime}}^{2}\ \vec{r}\,\,_{22^{\prime}}^{2}}\nonumber\\
&  +\frac{\ln\left(  \frac{\vec{r}_{12^{\prime}}^{\;2}\,\vec{r}_{21^{\prime}%
}^{\;2}}{\vec{r}_{11^{\prime}}^{\;2}\vec{r}_{22^{\prime}}^{\;2}}\right)
}{4\vec{r}\,\,_{11^{\prime}}^{2}\ \vec{r}\,\,_{22^{\prime}}^{2}}\left(
\frac{\vec{r}\,\,_{12}^{4}}{d}-\frac{\vec{r}\,\,_{12}^{2}}{\vec{r}%
\,\,_{1^{\prime}2^{\prime}}^{2}}\right)  +\frac{\ln\left(  \frac{\vec
{r}\,\,_{12}^{2}\ \vec{r}\,\,_{1^{\prime}2^{\prime}}^{2}}{\vec{r}%
\,\,_{11^{\prime}}^{2}\vec{r}\,\,_{22^{\prime}}^{2}}\right)  }{2\vec
{r}\,\,_{12^{\prime}}^{2}\vec{r}\,\,_{21^{\prime}}^{2}\ }\left(  \frac{\vec
{r}\,\,_{12}^{2}}{2\vec{r}\,\,_{1^{\prime}2^{\prime}}^{2}}+\frac{1}{2}%
-\frac{\vec{r}\,\,_{22^{\prime}}^{2}}{\vec{r}\,\,_{1^{\prime}2^{\prime}}^{2}%
}\right)  \nonumber\\
&  +\frac{\vec{r}\,\,_{21^{\prime}}^{2}\ln\left(  \frac{\vec{r}%
\,\,_{21^{\prime}}^{2}\vec{r}\,\,_{1^{\prime}2^{\prime}}^{2}}{\vec{r}%
\,\,_{12}^{2}\ \vec{r}\,\,_{11^{\prime}}^{2}}\right)  }{2\vec{r}%
\,\,_{11^{\prime}}^{2}\vec{r}\,\,_{22^{\prime}}^{2}\vec{r}\,\,_{1^{\prime
}2^{\prime}}^{2}}+\frac{\ln\ \left(  \frac{\vec{r}\,\,_{12}^{2}}{\vec
{r}\,\,_{1^{\prime}2^{\prime}}^{2}}\right)  }{4\vec{r}\,\,_{11^{\prime}}%
^{2}\vec{r}\,\,_{22^{\prime}}^{2}}+\frac{\ln\left(  \frac{\vec{r}%
\,\,_{22^{\prime}}^{2}}{\vec{r}\,\,_{12}^{2}}\right)  }{2\vec{r}%
\,\,_{11^{\prime}}^{2}\ \vec{r}\,\,_{12^{\prime}}^{2}}+\frac{\vec{r}%
\,\,_{12}^{2}\ln\ \left(  \frac{\vec{r}\,\,_{12}^{2}\vec{r}\,\,_{1^{\prime
}2^{\prime}}^{2}}{\vec{r}\,\,_{12^{\prime}}^{2}\ \vec{r}\,\,_{21^{\prime}}%
^{2}}\right)  }{4\vec{r}\,\,_{11^{\prime}}^{2}\vec{r}\,\,_{22^{\prime}}%
^{2}\vec{r}\,\,_{1^{\prime}2^{\prime}}^{2}}\nonumber\\
&  +\frac{\ln\ \left(  \frac{\vec{r}\,\,_{12}^{2}\vec{r}\,\,_{1^{\prime
}2^{\prime}}^{2}}{\vec{r}\,\,_{12^{\prime}}^{2}\vec{r}\,\,_{22^{\prime}}^{2}%
}\right)  }{2\vec{r}\,\,_{11^{\prime}}^{2}\ \vec{r}\,\,_{1^{\prime}2^{\prime}%
}^{2}}+\frac{\ln\left(  \frac{\vec{r}\,\,_{12}^{2}\vec{r}\,\,_{11^{\prime}%
}^{2}}{\vec{r}\,\,_{22^{\prime}}^{2}\ \vec{r}\,\,_{1^{\prime}2^{\prime}}^{2}%
}\right)  }{2\vec{r}\,\,_{12^{\prime}}^{2}\vec{r}\,\,_{1^{\prime}2^{\prime}%
}^{2}}+(1\leftrightarrow2,1^{\prime}\leftrightarrow2^{\prime}),\quad\quad
d=\vec{r}_{12^{\prime}}^{\;2}\vec{r}_{21^{\prime}}^{\;2}-\vec{r}_{11^{\prime}%
}^{\;2}\vec{r}_{22^{\prime}}^{\,\,2}.\label{gBFKL}%
\end{align}
Remind that the coefficients of $\delta(\vec{r}_{11^{\prime}})\delta
({r}_{22^{\prime}})$ in Eq.~(\ref{K_in_corrdinate_repr}) are written in the
integral form in order to make explicit cancellation of the ultraviolet
singularities of separate terms. Therefore one can take $g_{0}$ in various
forms (without change of the integral $\int d\vec{\rho}\,g_{0}(\vec{r}%
_{1},\vec{r}_{2};\rho)$). Here we change the form of $g_{0}$ in comparison
with our previous papers using the equalities
\begin{equation}
\int d\vec{\rho}\frac{\,\vec{r}_{12}^{\;2}}{\vec{r}_{1\rho}^{\;2}\vec
{r}_{2\rho}^{\;2}}\ln\left(  \frac{\vec{r}_{1\rho}^{\;2}}{\vec{r}_{12}^{\;2}%
}\right)  \ln\left(  \frac{\vec{r}_{2\rho}^{\;2}}{\vec{r}_{12}^{\;2}}\right)
=\int\frac{d\vec{\rho}\,}{\vec{r}_{2\rho}^{\;2}}\ln\left(  \frac{\vec
{r}_{1\rho}^{\;2}}{\vec{r}_{12}^{\;2}}\right)  \ln\left(  \frac{\vec{r}%
_{1\rho}^{\;2}}{\vec{r}_{2\rho}^{\;2}}\right)  =4\pi\zeta
(3).\label{representation for zeta in coordinate space}%
\end{equation}
In the colour dipole approach the gluon contribution was found in
Ref.~\cite{Balitsky:2008zz}. With account of the correction given in
Ref.~\cite{Balitsky:2009xg}, it gives
\begin{equation}
g_{0}^{BC}(\vec{r}_{1},\vec{r}_{2},\vec{\rho})=2\pi\zeta(3)\delta\left(
\vec{\rho}\right)  -g_{1}^{BC}(\vec{r}_{1},\vec{r}_{2},\vec{\rho
}),\label{g0BK}%
\end{equation}%
\[
g_{1}^{BC}(\vec{r}_{1},\vec{r}_{2};\vec{r}_{2}^{\;\prime})\ =\frac{11}{6}%
\frac{\vec{r}_{12}^{\;2}}{\vec{r}_{22^{\prime}}^{\;2}\vec{r}_{12^{\prime}%
}^{\;2}}\ln\left(  \frac{\vec{r}_{12}^{\;2}}{r_{\mu_{BC}}^{2}}\right)
+\frac{11}{6}\left(  \frac{1}{\vec{r}_{22^{\prime}}^{\,\,\,2}}-\frac{1}%
{\vec{r}_{12^{\prime}}^{\,\,\,2}}\right)  \ln\left(  \frac{\vec{r}%
_{22^{\prime}}^{\,\,\,2}}{\vec{r}_{12^{\prime}}^{\,\,\,2}}\right)
\]%
\begin{equation}
-\frac{\vec{r}_{12}^{\;2}}{\vec{r}_{22^{\prime}}^{\;2}\vec{r}_{12^{\prime}}%
}\ln\left(  \frac{\vec{r}_{12}^{\;2}}{\vec{r}_{22^{\prime}}}\right)
\ln\left(  \frac{\vec{r}_{12}^{\;2}}{\vec{r}_{12^{\prime}}}\right)
,\label{g1BK}%
\end{equation}
where
\begin{equation}
\ln r_{\mu_{BC}}^{2}=-\ln\mu^{2}-\frac{3}{11}\left(  \frac{67}{9}%
-2\zeta(2)\right)  ,
\end{equation}%
\[
g_{2}^{BC}(\vec{r}_{1},\vec{r}_{2};\vec{r}_{1}^{\;\prime},\vec{r}%
_{2}^{\;\prime})=\ln\left(  \frac{\vec{r}_{12^{\prime}}^{\,\,2}\vec
{r}_{21^{\prime}}^{\,\,2}}{\vec{r}_{11^{\prime}}^{\,\,2}\vec{r}_{22^{\prime}%
}^{\,\,2}}\right)  \left[  \frac{\vec{r}_{11^{\prime}}^{\,\,2}\vec
{r}_{22^{\prime}}^{\,\,2}+\vec{r}_{12^{\prime}}^{\;2}\vec{r}_{21^{\prime}%
}^{\;2}-4\vec{r}_{12}^{\,\,2}\vec{r}_{1^{\prime}2^{\prime}}^{\,\,2}}%
{2d\,\vec{r}_{1^{\prime}2^{\prime}}^{\,\,4}}\right.
\]%
\begin{equation}
\left.  +\frac{1}{4\vec{r}_{11^{\prime}}^{\,\,2}\vec{r}_{22^{\prime}}^{\,\,2}%
}\left(  \frac{\vec{r}_{12}^{\,\,4}}{d}-\frac{\vec{r}_{12}^{\,\,2}}{\vec
{r}_{1^{\prime}2^{\prime}}^{\,\,2}}\right)  +\frac{1}{4\vec{r}_{12^{\prime}%
}^{\,\,2}\vec{r}_{21^{\prime}}^{\,\,2}}\left(  \frac{\vec{r}_{12}^{\,\,4}}%
{d}+\frac{\vec{r}_{12}^{\,\,2}}{\vec{r}_{1^{\prime}2^{\prime}}^{\,\,2}%
}\right)  \right]  -\frac{1}{\vec{r}_{1^{\prime}2^{\prime}}^{\,\,4}%
}.\label{g2BK}%
\end{equation}

\section{Matching of the gluon parts}

In principle, if the kernels $\hat{\emph{K}}_{M}$ and $\hat{\mathcal{K}}_{BC}$
can be connected by the transformation (\ref{kernel_transforms}), one can
easily write a formal expression for the operator ${\hat{O}}$. Indeed, let us
denote $\hat{\emph{K}}_{M}-\hat{\mathcal{K}}_{BC}=\hat{{\Delta}}$, the
eigenstates of the Born kernel $\hat{\mathcal{K}}^{B}$ $|\mu\rangle$, and the
corresponding eigenvalues $\omega_{\mu}^{B}$. Then, if $\hat{{\Delta}}=\left[
\hat{\mathcal{K}}^{B},\hat{{O}}\right]  $, one has
\begin{equation}
\left(  \omega_{\mu^{\prime}}^{B}-\omega_{\mu}^{B}\right)  \langle\mu^{\prime
}|\hat{{O}}|\mu\rangle=\langle\mu^{\prime}|\hat{\Delta}|\mu\rangle.
\end{equation}
It can be seen from this equation that the operator ${\hat{O}}$ exists only if
the operator $\hat{\Delta}$ has zero matrix elements between states of equal
eigenvalues. If so, supposing that the states $|\mu\rangle$ form a complete
set, one has
\begin{equation}
\hat{{O}}=\sum_{\mu,\mu^{\prime}}\frac{|\mu^{\prime}\rangle\langle\mu^{\prime
}|\hat{\Delta}|\mu\rangle\langle\mu|}{\omega_{\mu^{\prime}}^{B}-\omega_{\mu
}^{B}}%
\end{equation}
and
\begin{equation}
\langle\vec{r}_{1},\vec{r}_{2}|\hat{O}|\vec{r}_{1}^{\;\prime},\vec{r}%
_{2}^{\;\prime}\rangle=\sum_{\mu,\mu^{\prime}}\frac{\langle\vec{r}_{1},\vec
{r}_{2}|\mu^{\prime}\rangle\langle\mu^{\prime}|\hat{\Delta}|\mu\rangle
\langle\mu|\vec{r}_{1}^{\;\prime},\vec{r}_{2}^{\;\prime}\rangle}{\omega
_{\mu^{\prime}}^{B}-\omega_{\mu}^{B}}.\label{forml O}%
\end{equation}
Since we know $\langle\vec{r}_{1},\vec{r}_{2}|\hat{\Delta}|\vec{r}%
_{1}^{\;\prime},\vec{r}_{2}^{\;\prime}\rangle$ from
Eqs.~(\ref{K_in_corrdinate_repr}), (\ref{g0BFKL})-(\ref{gBFKL}) and(\ref{g0BK}%
)-(\ref{g2BK}), we can find
\begin{equation}
\langle\mu^{\prime}|\hat{\Delta}|\mu\rangle=\int d\vec{r}_{1}d\vec{r}_{2}%
d\vec{r}_{1}^{\;\prime}d\vec{r}_{2}^{\;\prime}\langle\mu^{\prime}|\vec{r}%
_{1}^{\;\prime},\vec{r}_{2}^{\;\prime}\rangle\langle\vec{r}_{1}^{\;\prime
},\vec{r}_{2}^{\;\prime}|\hat{\Delta}|\vec{r}_{1},\vec{r}_{2}\rangle
\langle\vec{r}_{1},\vec{r}_{2}|\mu\rangle
\end{equation}
using the known eigenfunctions $\langle\vec{r}_{1},\vec{r}_{2}|\mu\rangle$ in
the coordinate space \cite{Lipatov:1985uk} and then $\langle\vec{r}_{1}%
,\vec{r}_{2}|\hat{O}|\vec{r}_{1}^{\;\prime},\vec{r}_{2}^{\;\prime}\rangle$
using Eq.~(\ref{forml O}). However, it is rather difficult because of
complexity of the eigenfunctions and the corresponding eigenvalues. In fact,
we did not do it, but we have guessed the operator $\hat{O}$:
\[
\langle\vec{q}_{1},\vec{q}_{2}|{\hat{O}}|\vec{q}_{1}^{\,\,\prime},\vec{q}%
_{2}^{\,\,\prime}\rangle=-\delta(\vec{q}_{11^{\prime}}+\vec{q}_{22^{\prime}%
})\frac{\mathcal{K}_{r}^{B}(\vec{q}_{1},\vec{q}_{1}^{\;\prime};\vec{q})}%
{2\vec{q}_{1}^{\,\,2}\vec{q}_{2}^{\,\,2}}\ln\vec{q}_{11^{\prime}}^{\,\,2}%
\]%
\begin{equation}
+\frac{\alpha_{s}N_{c}}{4\pi^{2}}\,\delta(\vec{q}_{22^{\prime}})\delta\left(
\vec{q}_{11^{\prime}}\right)  \int d^{2+2\epsilon}k\ln\vec{k}^{\,\,2}\left(
\frac{2}{\vec{k}^{\,\,2}}-\frac{\vec{k}(\vec{k}-\vec{q}_{1})}{\vec{k}%
^{\,\,2}(\vec{k}-\vec{q}_{1})^{2}}-\frac{\vec{k}(\vec{k}-\vec{q}_{2})}{\vec
{k}^{\,\,2}(\vec{k}-\vec{q}_{2})^{2}}\right)  .\label{O}%
\end{equation}
Let us show that the transformation defined by Eqs.~(\ref{kernel_transforms})
and (\ref{O}), being applied to the kernel $\hat{\emph{K}}_{M}$, converts the
functions $g_{i}$ given by Eqs.~(\ref{K_in_corrdinate_repr}) and
(\ref{g0BFKL})-(\ref{gBFKL}) into the functions $g_{i}^{BC}$ (\ref{g0BK}%
)-(\ref{g2BK}).

In the momentum space, we have for the commutator $[\mathcal{\hat{K}\,}%
^{B},\hat{O}]$
\[
\langle\vec{q}_{1},\vec{q}_{2}|[\mathcal{\hat{K}\,}^{B},\hat{O}]\mathcal{\,}%
|\vec{q}_{1}^{\,\,\prime},\vec{q}_{2}^{\,\,\prime}\rangle=\delta(\vec
{q}_{11^{\prime}}+\vec{q}_{22^{\prime}})\left[  \int d\vec{k}\,\,\frac
{\mathcal{K}_{r}^{B}(\vec{q}_{1},\vec{q}_{1}^{\;}-\vec{k};\vec{q})}{2\vec
{q}_{1}^{\,\,2}\vec{q}_{2}^{\,\,2}}\frac{\mathcal{K}_{r}^{B}(\vec{q}_{1}%
-\vec{k},\vec{q}_{1}^{\;\prime};\vec{q})}{(\vec{k}-\vec{q}_{1})^{2}(\vec
{k}+\vec{q}_{2})^{2}}\,\ln\frac{\vec{k}^{\,\,2}}{(\vec{k}-\vec{q}_{11^{\prime
}})^{2}}\right.
\]%
\begin{equation}
\left.  +\frac{\alpha_{s}N_{c}}{4\pi^{2}}\frac{\mathcal{K}_{r}^{B}(\vec{q}%
_{1},\vec{q}_{1}^{\;\prime};\vec{q})}{\vec{q}_{1}^{\,\,2}\vec{q}_{2}^{\,\,2}%
}\int d\vec{k}\left(  \frac{\vec{k}(\vec{k}-\vec{q}_{1}^{\,\,\prime})}{\vec
{k}^{\,\,2}(\vec{k}-\vec{q}_{1}^{\,\,\prime})^{2}}-\frac{\vec{k}(\vec{k}%
-\vec{q}_{1})}{\vec{k}^{\,\,2}(\vec{k}-\vec{q}_{1})^{2}}+(1\leftrightarrow
2,1^{\prime}\leftrightarrow2^{\prime})\right)  \ln\frac{\vec{q}_{11^{\prime}%
}^{\,\,2}}{\vec{k}^{\,\,2}}\right]  .\label{[KO]}%
\end{equation}
Since all integrals here and below are convergent, we put $\epsilon
=\allowbreak0$ henceforth. The first line in Eq.~(\ref{[KO]}) is equal to the
doubled contribution $\langle\vec{q}_{1},\vec{q}_{2}|\mathcal{\hat{K}}%
_{s2}|\vec{q}_{1}^{\,\,\prime},\vec{q}_{2}^{\,\,\prime}\rangle$ to the BFKL
kernel, defined in Ref.~\cite{Fadin:2007de}. Denoting the remaining terms in
Eq.~(\ref{[KO]}) as $\langle\vec{q}_{1},\vec{q}_{2}|\mathcal{\hat{V}}|\vec
{q}_{1}^{\,\,\prime},\vec{q}_{2}^{\,\,\prime}\rangle$, after integration we
obtain
\[
\langle\vec{q}_{1},\vec{q}_{2}|\mathcal{\hat{V}}|\vec{q}_{1}^{\,\,\prime}%
,\vec{q}_{2}^{\,\,\prime}\rangle=\delta(\vec{q}_{11^{\prime}}+\vec
{q}_{22^{\prime}})\frac{\left(  \alpha_{s}N_{c}\right)  ^{2}}{8\pi^{3}}\left(
\frac{1}{\vec{k}^{\,\,2}}+\frac{(\vec{q}_{2}\,\vec{k})}{\vec{q}_{2}%
^{\,\,2}\vec{k}^{\,\,2}}-\frac{(\vec{q}_{1}\,\vec{k})}{\vec{q}_{1}^{\,\,2}%
\vec{k}^{\,\,2}}-\frac{\left(  \vec{q}_{1}\,\vec{q}_{2}\right)  }{\vec{q}%
_{1}^{\,\,2}\vec{q}_{2}^{\,\,2}}\right)
\]%
\begin{equation}
\times\left(  \ln^{2}\frac{\vec{q}_{1}^{\,\,\prime\,2}}{\vec{k}^{\,\,2}}%
+\ln^{2}\frac{\vec{q}_{2}^{\,\,\prime2}}{\vec{k}^{\,\,2}}-\ln^{2}\frac{\vec
{q}_{1}^{\,\,2}}{\vec{k}^{\,\,2}}-\ln^{2}\frac{\vec{q}_{2}^{\,\,2}}{\vec
{k}^{\,\,2}}\right)  ,\label{V}%
\end{equation}
where $\vec{k}=\vec{q}_{11^{\prime}}.$

The M\"{o}bius form of $\mathcal{\hat{K}}_{s2}$ was found in
Ref.~\cite{Fadin:2007de}. To obtain such form for the operator $\mathcal{\hat
{V}}$ we have to transform Eq.~(\ref{V}) into the coordinate space. It can be
done using the following integrals:
\begin{equation}
\int\frac{d\vec{k}}{2\pi}e^{i\vec{k}\,\vec{r}}\frac{\vec{k}}{\vec{k}^{\,2}%
}=\frac{i\vec{r}}{\vec{r}^{\,2}},
\end{equation}%
\begin{equation}
\int\frac{d\vec{q}}{2\pi}\int\frac{d\vec{k}}{2\pi}e^{i[\vec{q}\,\vec{r}%
+\vec{k}\,\vec{\rho}]}\frac{1}{\vec{q}^{\,2}}\ln\frac{(\vec{q}-\vec{k})^{2}%
}{\vec{k}^{\,2}}\ln\frac{(\vec{q}-\vec{k})^{2}}{\vec{q}^{\,2}}=\frac{1}%
{\vec{\rho}^{\,2}}\ln\left(  \frac{\left(  \vec{r}+\vec{\rho}\right)  ^{2}%
}{\vec{r}^{\,2}}\right)  \ln\left(  \frac{\left(  \vec{r}+\vec{\rho}\right)
^{2}}{\vec{\rho}^{\,2}}\right)  ,\label{int 1/q ln(q+k)/q ln(q+k)/k}%
\end{equation}%
\begin{equation}
\int\frac{d\vec{q}}{2\pi}\int\frac{d\vec{k}}{2\pi}e^{i[\vec{q}\,\vec{r}%
+\vec{k}\,\vec{\rho}]}\frac{1}{\vec{q}^{\,2}}\ln^{2}\frac{(\vec{q}-\vec
{k})^{2}}{\vec{k}^{\,2}}=\frac{1}{\vec{\rho}^{\,2}}\ln^{2}\left(
\frac{\left(  \vec{r}+\vec{\rho}\right)  ^{2}}{\vec{r}^{\,2}}\right)  ,
\end{equation}%
\begin{equation}
\int\frac{d\vec{q}}{2\pi}\int\frac{d\vec{k}}{2\pi}e^{i[\vec{q}\,\vec{r}%
+\vec{k}\,\vec{\rho}]}\frac{(\vec{q}\,\vec{k})}{\vec{q}^{\,2}\vec{k}^{\,2}}%
\ln^{2}\frac{(\vec{k}+\vec{q})^{2}}{\vec{q}^{\,2}}=-\frac{\left(  \vec
{r}\,\vec{\rho}\right)  }{\vec{r}^{\,2}\vec{\rho}^{\,2}}\ln^{2}\left(
\frac{(\vec{\rho}-\vec{r})^{2}}{\vec{\rho}^{\,2}}\right)  ,
\end{equation}%
\begin{equation}
\int\frac{d\vec{q}}{2\pi}\int\frac{d\vec{k}}{2\pi}e^{i[\vec{q}\,\vec{r}%
+\vec{k}\,\vec{\rho}]}\frac{(\vec{q}\,\vec{k})}{\vec{q}^{\,2}\vec{k}^{\,2}}%
\ln^{2}\frac{\vec{k}^{\,2}}{\vec{q}^{\,2}}=-\frac{\left(  \vec{r}\,\vec{\rho
}\right)  }{\vec{r}^{\,2}\vec{\rho}^{\,2}}\ln^{2}\left(  \frac{\vec{\rho
}^{\,2}}{\vec{r}^{\,2}}\right)  ,
\end{equation}%
\begin{equation}
\int\frac{d\vec{q}_{1}}{2\pi}\int\frac{d\vec{q}_{2}}{2\pi}\int\frac{d\vec{k}%
}{2\pi}e^{i[\vec{q}_{1}\,\vec{r}_{1}+\vec{q}_{2}\,\vec{r}_{2}+\vec{k}%
\,\vec{\rho}]}\frac{(\vec{q}_{1}\,\vec{q}_{2})}{\vec{q}_{1}^{\;2}\vec{q}%
_{2}^{\;2}}\ln^{2}\frac{\vec{q}_{1}^{\;2}}{\vec{k}^{\,2}}=\frac{4(\vec{r}%
_{1}\,\vec{r}_{2})}{\vec{r}_{1}^{\,2}\vec{r}_{2}^{\,2}\vec{\rho}^{\,2}}%
\ln\left(  \frac{\vec{r}_{1}^{\,2}}{\vec{\rho}^{\,2}}\right)
.\label{int (q1q2)/q1q2 ln q1/k}%
\end{equation}
The result is
\[
\frac{8\pi^{4}}{\left(  \alpha_{s}N_{c}\right)  ^{\,\,2}}\langle\vec{r}%
_{1}\vec{r}_{2}|\mathcal{\hat{V}}|\vec{r}_{1}^{\;\prime}\vec{r}_{2}^{\;\prime
}\rangle=v_{M}(\vec{r}_{1},\vec{r}_{2};\vec{r}_{1}^{\;\prime},\vec{r}%
_{2}^{\;\prime})
\]%
\begin{equation}
+\left[  \frac{\vec{r}_{11^{\prime}}^{\,\,2}-\vec{r}_{1^{\prime}2^{\prime}%
}^{\,\,2}}{\vec{r}_{11^{\prime}}^{\,\,2}\vec{r}_{12^{\prime}}^{\,\,2}\vec
{r}_{1^{\prime}2^{\prime}}^{\,\,2}}\ln\left(  \frac{\vec{r}_{11^{\prime}%
}^{\,\,2}}{\vec{r}_{12^{\prime}}^{\,\,2}}\right)  +\frac{1}{\vec
{r}_{11^{\prime}}^{\,\,2}\vec{r}_{1^{\prime}2^{\prime}}^{\,\,2}}\ln\left(
\frac{\vec{r}_{11^{\prime}}^{\,\,2}\vec{r}_{12^{\prime}}^{\,\,2}}{\vec
{r}_{1^{\prime}2^{\prime}}^{\,\,4}}\right)  +(1\leftrightarrow2,1^{\prime
}\leftrightarrow2^{\prime})\right]  ,\label{V coordinate}%
\end{equation}
where%
\[
v_{M}(\vec{r}_{1},\vec{r}_{2};\vec{r}_{1}^{\;\prime},\vec{r}_{2}^{\;\prime
})=\pi\delta(\vec{r}_{22^{\prime}})\frac{\vec{r}_{12}^{\,\,2}+\vec
{r}_{21^{\prime}}^{\,\,2}-\vec{r}_{11^{\prime}}^{\,\,2}}{2\vec{r}_{11^{\prime
}}^{\,\,2}\vec{r}_{21^{\prime}}^{\,\,2}}\ln\left(  \frac{\vec{r}_{21^{\prime}%
}^{\,\,2}}{\vec{r}_{12}^{\,\,2}}\right)  \ln\left(  \frac{\vec{r}_{11^{\prime
}}^{\,\,4}}{\vec{r}_{12}^{\,\,2}\vec{r}_{21^{\prime}}^{\,\,2}}\right)
\]%
\[
-\vec{r}_{12}^{\,\,2}\left(  \frac{\ln\left(  \frac{\vec{r}_{12^{\prime}%
}^{\,\,2}}{\vec{r}_{11^{\prime}}^{\,\,2}}\right)  }{\vec{r}_{11^{\prime}%
}^{\,\,2}\vec{r}_{12^{\prime}}^{\,\,2}\vec{r}_{22^{\prime}}^{\,\,2}}+\frac
{\ln\left(  \frac{\vec{r}_{22^{\prime}}^{\,\,2}}{\vec{r}_{21^{\prime}}%
^{\,\,2}}\right)  }{\vec{r}_{11^{\prime}}^{\,\,2}\vec{r}_{21^{\prime}}%
^{\,\,2}\vec{r}_{1^{\prime}2^{\prime}}^{\,\,2}}+\frac{\ln\left(  \frac{\vec
{r}_{12^{\prime}}^{\,\,2}}{\vec{r}_{1^{\prime}2^{\prime}}^{\,\,2}}\right)
}{\vec{r}_{11^{\prime}}^{\,\,2}\vec{r}_{22^{\prime}}^{\,\,2}\vec{r}%
_{1^{\prime}2^{\prime}}^{\,\,2}}\right)
\]%
\begin{equation}
-\frac{\vec{r}_{22^{\prime}}^{\,\,2}-\vec{r}_{12^{\prime}}^{\,\,2}}{\vec
{r}_{11^{\prime}}^{\,\,2}\vec{r}_{22^{\prime}}^{\,\,2}\vec{r}_{1^{\prime
}2^{\prime}}^{\,\,2}}\ln\left(  \frac{\vec{r}_{12^{\prime}}^{\,\,2}\vec
{r}_{21^{\prime}}^{\,\,2}}{\vec{r}_{22^{\prime}}^{\,\,2}\vec{r}_{1^{\prime
}2^{\prime}}^{\,\,2}}\right)  +(1\leftrightarrow2,1^{\prime}\leftrightarrow
2^{\prime}).\label{V_m}%
\end{equation}
The terms in the square brackets in Eq.~(\ref{V coordinate}) do not depend
either on $\vec{r}_{1}$ or on $\vec{r}_{2}$ and therefore they are omitted in
the M\"{o}bius form. As for $v_{M}(\vec{r}_{1},\vec{r}_{2};\vec{r}%
_{1}^{\;\prime},\vec{r}_{2}^{\;\prime})$, it turns into zero at $\vec{r}%
_{1}=\vec{r}_{2}$, so that it satisfies the requirements for the M\"{o}bius
forms. Using the M\"{o}bius form of $\mathcal{\hat{K}}_{s2}$
\cite{Fadin:2007de} (see Eqs.~(44), (54), (55), (67) and (69) there) and
Eq.~(\ref{V_m}), we obtain
\[
\frac{8\pi^{4}}{\left(  \alpha_{s}N_{c}\right)  ^{\,\,2}}\langle\vec{r}%
_{1}\vec{r}_{2}|[\mathcal{\hat{K}\,}^{B},\hat{O}]_{M}|\vec{r}_{1}^{\;\prime
}\vec{r}_{2}^{\;\prime}\rangle=\pi\delta\left(  \vec{r}_{22^{\prime}}\right)
\left[  \frac{1}{\vec{r}_{11^{\prime}}^{\,\,2}}\ln\left(  \frac{\vec
{r}_{21^{\prime}}^{\,\,2}}{\vec{r}_{12}^{\,\,2}}\right)  \ln\left(  \frac
{\vec{r}_{11^{\prime}}^{\,\,2}}{\vec{r}_{21^{\prime}}^{\,\,2}}\right)
-\frac{\vec{r}_{12}^{\,\,2}}{\vec{r}_{11^{\prime}}^{\,\,2}\vec{r}_{21^{\prime
}}^{\,\,2}}\ln\left(  \frac{\vec{r}_{11^{\prime}}^{\,\,2}}{\vec{r}%
_{12}^{\,\,2}}\right)  \ln\left(  \frac{\vec{r}_{12}^{\,\,2}}{\vec
{r}_{21^{\prime}}^{\,\,2}}\right)  \right]
\]%
\[
-\vec{r}_{12}^{\,\,2}\left(  \frac{\ln\left(  \frac{\vec{r}_{12^{\prime}%
}^{\,\,2}\vec{r}_{21^{\prime}}^{\,\,2}}{\vec{r}_{12}^{\,\,2}\vec{r}%
_{1^{\prime}2^{\prime}}^{\,\,2}}\right)  }{2\vec{r}_{11^{\prime}}^{\,\,2}%
\vec{r}_{22^{\prime}}^{\,\,2}\vec{r}_{1^{\prime}2^{\prime}}^{\,\,2}}+\frac
{\ln\left(  \frac{\vec{r}_{1^{\prime}2^{\prime}}^{\,\,2}}{\vec{r}_{22^{\prime
}}^{\,\,2}}\right)  }{\vec{r}_{11^{\prime}}^{\,\,2}\vec{r}_{21^{\prime}%
}^{\,\,2}\vec{r}_{22^{\prime}}^{\,\,2}}+\frac{\ln\left(  \frac{\vec
{r}_{11^{\prime}}^{\,\,2}\vec{r}_{22^{\prime}}^{\,\,2}}{\vec{r}_{12}%
^{\,\,2}\vec{r}_{1^{\prime}2^{\prime}}^{\,\,2}}\right)  }{2\vec{r}%
_{12^{\prime}}^{\,\,2}\vec{r}_{21^{\prime}}^{\,\,2}\vec{r}_{1^{\prime
}2^{\prime}}^{\,\,2}}\right)  -\frac{\ln\left(  \frac{\vec{r}_{22^{\prime}%
}^{\,\,2}\vec{r}_{1^{\prime}2^{\prime}}^{\,\,2}}{\vec{r}_{12}^{\,\,2}\vec
{r}_{12^{\prime}}^{\,\,2}}\right)  }{\vec{r}_{12^{\prime}}^{\,\,2}\vec
{r}_{1^{\prime}2^{\prime}}^{\,\,2}}-\frac{\ln\left(  \frac{\vec{r}%
_{12}^{\,\,2}\vec{r}_{11^{\prime}}^{\,\,2}}{\vec{r}_{21^{\prime}}^{\,\,2}%
\vec{r}_{1^{\prime}2^{\prime}}^{\,\,2}}\right)  \vec{r}_{21^{\prime}}^{\,\,2}%
}{\vec{r}_{11^{\prime}}^{\,\,2}\vec{r}_{22^{\prime}}^{\,\,2}\vec{r}%
_{1^{\prime}2^{\prime}}^{\,\,2}}%
\]%
\begin{equation}
-\frac{\ln\left(  \frac{\vec{r}_{12}^{\,\,2}\vec{r}_{12^{\prime}}^{\,\,2}%
}{\vec{r}_{11^{\prime}}^{\,\,2}\vec{r}_{22^{\prime}}^{\,\,2}}\right)  }%
{\vec{r}_{11^{\prime}}^{\,\,2}\vec{r}_{12^{\prime}}^{\,\,2}}-\frac{\ln\left(
\frac{\vec{r}_{1^{\prime}2^{\prime}}^{\,\,2}}{\vec{r}_{12}^{\,\,2}}\right)
}{2\vec{r}_{11^{\prime}}^{\,\,2}\vec{r}_{22^{\prime}}^{\,\,2}}-\frac
{\ln\left(  \frac{\vec{r}_{11^{\prime}}^{\,\,2}\vec{r}_{22^{\prime}}^{\,\,2}%
}{\vec{r}_{12}^{\,\,2}\vec{r}_{1^{\prime}2^{\prime}}^{\,\,2}}\right)  }%
{2\vec{r}_{12^{\prime}}^{\,\,2}\vec{r}_{21^{\prime}}^{\,\,2}}-\frac{\ln\left(
\frac{\vec{r}_{12}^{\,\,2}\vec{r}_{1^{\prime}2^{\prime}}^{\,\,2}}{\vec
{r}_{11^{\prime}}^{\,\,2}\vec{r}_{22^{\prime}}^{\,\,2}}\right)  \vec
{r}_{22^{\prime}}^{\,\,2}}{\vec{r}_{12^{\prime}}^{\,\,2}\vec{r}_{21^{\prime}%
}^{\,\,2}\vec{r}_{1^{\prime}2^{\prime}}^{\,\,2}}-\frac{\ln\left(  \frac
{\vec{r}_{11^{\prime}}^{\,\,2}\vec{r}_{22^{\prime}}^{\,\,2}}{\vec{r}%
_{12}^{\,\,2}\vec{r}_{1^{\prime}2^{\prime}}^{\,\,2}}\right)  }{\vec
{r}_{11^{\prime}}^{\,\,2}\vec{r}_{1^{\prime}2^{\prime}}^{\,\,2}}%
+(1\leftrightarrow2,1^{\prime}\leftrightarrow2^{\prime}).
\end{equation}
>From the definition (\ref{K_in_corrdinate_repr}) it follows that the
transformation $\hat{\emph{K}}_{M}\rightarrow\ \hat{{\mathcal{K}}}%
_{M}-[\mathcal{\hat{K}\,}^{B},\hat{O}]_{M}$ leaves $g_{0}$ untouched and
changes only $g_{1,2}$. Using Eqs.~(\ref{g1BFKL}), (\ref{gBFKL}) we get
\[
g_{1}(\vec{r}_{1},\vec{r}_{2};\vec{r}_{2}^{\;\prime})\ \rightarrow g_{1}%
^{T}(\vec{r}_{1},\vec{r}_{2};\vec{r}_{2}^{\;\prime})\ =\frac{11}{6}\frac
{\vec{r}_{12}^{\;2}}{\vec{r}_{22^{\prime}}^{\;2}\vec{r}_{12^{\prime}}^{\;2}%
}\ln\left(  \frac{\vec{r}_{12}^{\;2}}{r_{\mu}^{2}}\right)  +\frac{11}%
{6}\left(  \frac{1}{\vec{r}_{22^{\prime}}^{\,\,\,2}}-\frac{1}{\vec
{r}_{12^{\prime}}^{\,\,\,2}}\right)  \ln\left(  \frac{\vec{r}_{22^{\prime}%
}^{\,\,\,2}}{\vec{r}_{12^{\prime}}^{\,\,\,2}}\right)
\]%
\begin{equation}
-\frac{\vec{r}_{12}^{\;2}}{\,\vec{r}_{22^{\prime}}^{\;2}\vec{r}_{12^{\prime}%
}^{\;2}}\ln\left(  \frac{\vec{r}_{12}^{\;2}}{\vec{r}_{22^{\prime}}^{\;2}%
}\right)  \ln\left(  \frac{\vec{r}_{12}^{\;2}}{\vec{r}_{12^{\prime}}^{\;2}%
}\right)  ,\label{g1T}%
\end{equation}
and
\[
g_{2}(\vec{r}_{1},\vec{r}_{2};\vec{r}_{1}^{\;\prime},\vec{r}_{2}^{\;\prime
})\rightarrow g_{2}^{T}(\vec{r}_{1},\vec{r}_{2};\vec{r}_{1}^{\;\prime},\vec
{r}_{2}^{\;\prime})=g_{2}^{T(s)}(\vec{r}_{1},\vec{r}_{2};\vec{r}_{1}%
^{\;\prime},\vec{r}_{2}^{\;\prime})+\left[  \left(  \frac{(\vec{r}%
_{12^{\prime}}\,\,\vec{r}_{11^{\prime}})}{\vec{r}_{11^{\prime}}^{\,\,2}\vec
{r}_{12^{\prime}}^{\,\,2}\vec{r}_{1^{\prime}2^{\prime}}^{\,\,2}}+\frac
{1}{2\vec{r}_{1^{\prime}2^{\prime}}^{\,\,4}}\right.  \right.
\]%
\begin{equation}
\left.  \left.  +\frac{\vec{r}_{12}^{\,\,2}}{4\vec{r}_{1^{\prime}2^{\prime}%
}^{\,\,2}}\left(  \frac{1}{\vec{r}_{11^{\prime}}^{\,\,2}\vec{r}_{22^{\prime}%
}^{\,\,2}}+\frac{1}{\vec{r}_{12^{\prime}}^{\,\,2}\vec{r}_{21^{\prime}}%
^{\,\,2}}\right)  -\frac{\vec{r}_{12}^{\,\,2}}{4\vec{r}_{11^{\prime}}%
^{\,\,2}\vec{r}_{22^{\prime}}^{\,\,2}\vec{r}_{12^{\prime}}^{\,\,2}\vec
{r}_{21^{\prime}}^{\,\,2}}\right)  \ln\left(  \frac{\vec{r}_{11^{\prime}%
}^{\,\,2}}{\vec{r}_{12^{\prime}}^{\,\,2}}\right)  +(1\leftrightarrow
2,1^{\prime}\leftrightarrow2^{\prime})\right]  ,\label{g2T}%
\end{equation}
where
\begin{equation}
g_{2}^{T(s)}(\vec{r}_{1},\vec{r}_{2};\vec{r}_{1}^{\;\prime},\vec{r}%
_{2}^{\;\prime})=g_{2}^{BC}(\vec{r}_{1},\vec{r}_{2};\vec{r}_{1}^{\;\prime
},\vec{r}_{2}^{\;\prime})\label{g2Ts}%
\end{equation}
is symmetric with respect to the replacement ($\vec{r}_{1}^{\;\prime
}\leftrightarrow\vec{r}_{2}^{\;\prime}$) (as well as with respect to the
replacement ($\vec{r}_{1}\leftrightarrow\vec{r}_{2}$)), and the terms in the
square brackets are antisymmetric, so that they can be omitted with account of
the symmetry of impact factors. There is also another reason for omitting the
first term: it does not depend on $\vec{r}_{2}$. In the following we will
assume that this term is omitted. Note that using
Eq.~(\ref{representation for zeta in coordinate space}) we can rewrite
$g_{0}^{T}=g_{0}$ in the form
\begin{equation}
g_{0}^{T}(\vec{r}_{1},\vec{r}_{2},\vec{\rho})=2\pi\zeta(3)\delta\left(
\vec{\rho}\right)  -g_{1}^{T}(\vec{r}_{1},\vec{r}_{2},\vec{\rho}).\label{g0T}%
\end{equation}
Comparing Eqs.~(\ref{g1T}) and (\ref{g0T}) with Eqs.~(\ref{g1BK}) and
(\ref{g0BK}) we see that up to the normalization points the functions
$g_{1}^{T}$ and $g_{0}^{T}$ coincide with $g_{1}^{BC}$ and $g_{0}^{BC}$
correspondingly. Therefore, we conclude that the symmetrized gluon part of the
M\"{o}bius form of the kernel can be written as
\begin{equation}
\hat{\emph{K}}-[{\hat{\emph{K}}\,}^{B},\hat{O}]=\hat{\mathcal{K}}+\frac
{\alpha_{s}}{8\pi}\beta_{0}[\hat{\mathcal{K}}^{B},\ln\left(  \hat{\vec{q}}%
_{1}^{\;2}\hat{\vec{q}}_{2}^{\;2}\right)  ]-[{\hat{\mathcal{K}}\,}^{B},\hat
{O}],\label{matching kernel}%
\end{equation}
where $\beta_{0}$ is the first coefficient of the beta-function, the kernel
$\mathcal{K}$ is defined in Eq.~(\ref{K_momentum_repr_definition}) and the
operator $\hat{O}$ in Eq.~(\ref{O}). It coincides (up to the difference in the
renormalizations) with the gluon part of the kernel of the colour dipole
approach found in Ref.~\cite{Balitsky:2008zz} (with account of the correction
given in Ref.~\cite{Balitsky:2009xg}).

Since the M\"{o}bius form of the quark part of $\hat{\emph{K}}$ coincides with
the quark part of the linearized BK kernel \cite{Kovchegov:2006vj,
Balitsky:2006wa} calculated in the colour dipole model, and only gluons
contribute to $[{\hat{\mathcal{K}}\,}^{B},\hat{O}]$ (see Eq.~(\ref{[KO]})), it
means that the the symmetrized M\"{o}bius form of the kernel
(\ref{matching kernel}) coincides with the kernel of the colour dipole model
(up to the difference in the renormalization scales). Thus, the discrepancy
between the BFKL and the colour dipole approaches is completely removed.

\section{Transformation to the quasi-conformal shape}

As can be seen from the representation (\ref{K_in_corrdinate_repr}) and from
the explicit expressions for $g_{i}^{T}$, given by Eqs.~(\ref{g1T}%
)-(\ref{g0T}), (\ref{g2BK}), the conformal invariance of the M\"{o}bius form
of the kernel (\ref{matching kernel}) is violated not only by the terms
related to renormalization. However, from the results of
Ref.~\cite{Balitsky:2009xg} it is clear that we can transform the form
(\ref{matching kernel}) to the quasi-conformal kernel. Indeed, the
transformation from the usual kernel for the evolution of colour dipoles to
the kernel for the evolution of the \textquotedblleft composite dipole
operators" used in Ref.~\cite{Balitsky:2009xg} has the same nature as the
transformation (\ref{kernel_transforms}). Let us show that the transformation
$\hat{\emph{K}}\rightarrow\hat{\mathcal{K}}^{QC}=\hat{\emph{K}}-[\hat
{\mathcal{K}}^{B},O_{1}]$, where
\begin{equation}
\langle\vec{r}_{1}\vec{r}_{2}|\hat{{O}}_{1M}|\vec{r}_{1}^{\;\prime}\vec{r}%
_{2}^{\;\prime}\rangle=\frac{\alpha_{s}(\mu)N_{c}}{4\pi^{2}}\int d\vec{\rho
}\frac{\vec{r}_{12}{}^{2}}{\vec{r}_{1\rho}^{\,\,2}\vec{r}_{2\rho}^{\,\,2}}%
\ln\left(  \frac{\vec{r}_{12}{}^{2}}{\vec{r}_{1\rho}^{\,\,2}\vec{r}_{2\rho
}^{\,\,2}}\right)  \Biggl[\delta(\vec{r}_{11^{\prime}})\delta(\vec
{r}_{2^{\prime}\rho})+\delta(\vec{r}_{1^{\prime}\rho})\delta(\vec
{r}_{22^{\prime}})-\delta(\vec{r}_{11^{\prime}})\delta({r}_{22^{\prime}%
})\Biggr]~,\label{O1}%
\end{equation}
eliminates the nonconformal terms in $\hat{\emph{K}}_{M}$ which are not
proportional to the $\beta$--function.

Indeed, with the help of the integrals from Appendix A of
Ref.~\cite{Fadin:2009za}, for the commutator of this operator with the Born
part of the M\"{o}bius kernel (\ref{K_in_corrdinate_repr}) we obtain
\[
\langle\vec{r}_{1}\vec{r}_{2}|[\hat{\mathcal{K}}_{M}^{B}\,\,,\hat{{O}}%
_{1M}]|\vec{r}_{1}^{\;\prime}\vec{r}_{2}^{\;\prime}\rangle=-\frac{\alpha
_{s}^{2}(\mu)N_{c}^{2}}{4\pi^{4}}\left[  \frac{\vec{r}_{12}^{\;2}}{\vec
{r}\,\,_{11^{\prime}}^{2}\vec{r}\,\,_{22^{\prime}}^{2}\vec{r}_{1^{\prime
}2^{\prime}}^{\;2}}\ln\left(  \frac{\vec{r}_{12}^{\;2}\vec{r}_{1^{\prime
}2^{\prime}}^{\;2}}{\vec{r}_{12^{\prime}}^{\;2}\,\vec{r}_{21^{\prime}}^{\;2}%
}\right)  \right.
\]%
\begin{equation}
+\left.  \pi\delta(\vec{r}_{11^{\prime}})\frac{\vec{r}_{12}^{\;2}}{\,\vec
{r}_{22^{\prime}}^{\;2}\vec{r}_{12^{\prime}}^{\;2}}\ln\left(  \frac{\vec
{r}_{12}^{\;2}}{\vec{r}_{22^{\prime}}^{\;2}}\right)  \ln\left(  \frac{\vec
{r}_{12}^{\;2}}{\vec{r}_{12^{\prime}}^{\;2}}\right)  +\pi\delta({r}%
_{22^{\prime}})\frac{\vec{r}_{12}^{\;2}}{\,\vec{r}_{11^{\prime}}^{\;2}\vec
{r}_{21^{\prime}}^{\;2}}\ln\left(  \frac{\vec{r}_{12}^{\;2}}{\vec
{r}_{11^{\prime}}^{\;2}}\right)  \ln\left(  \frac{\vec{r}_{12}^{\;2}}{\vec
{r}_{21^{\prime}}^{\;2}}\right)  \right]  .
\end{equation}
Then, using the functions $g_{i}^{T}$ (\ref{g1T})--(\ref{g0T}) for the kernel
$\hat{\emph{K}}$, we obtain the functions $g_{i}^{QC}$ for the kernel
$\hat{\mathcal{K}}^{QC}= \hat{\emph{K}}-[\hat{\mathcal{K}}^{B}, O_{1}]$:
\begin{equation}
g^{QC}_{0}(\vec{r}_{1},\vec{r}_{2};\vec{\rho})=6\pi\zeta\left(  3\right)
\delta\left(  \vec{\rho}\right)  -g(\vec{r}_{1},\vec{r}_{2};\vec{\rho})~,
\label{gluon_g0}%
\end{equation}%
\begin{equation}
g^{QC}_{1}(\vec{r}_{1},\vec{r}_{2};\vec{r}_{2}^{\;\prime})\ =\frac{11}{6}%
\frac{\vec{r}_{12}^{\;2}}{\vec{r}_{22^{\prime}}^{\;2}\vec{r}_{12^{\prime}%
}^{\;2}}\ln\left(  \frac{\vec{r}_{12}^{\;2}}{r_{\mu}^{2}}\right)  +\frac
{11}{6}\left(  \frac{1}{\vec{r}_{22^{\prime}}^{\,\,\,2}}-\frac{1}{\vec
{r}_{12^{\prime}}^{\,\,\,2}}\right)  \ln\left(  \frac{\vec{r}_{22^{\prime}%
}^{\,\,\,2}}{\vec{r}_{12^{\prime}}^{\,\,\,2}}\right)  ,
\end{equation}
where $\ln r_{\mu}^{2}$ is defined in Eq.~(\ref{rmu}), and%
\begin{equation}
g^{QC}_{2}(\vec{r}_{1},\vec{r}_{2};\vec{r}_{1}^{\;\prime},\vec{r}%
_{2}^{\;\prime})=\frac{1}{\vec{r}_{1^{\prime}2^{\prime}}^{\,\,4}}\left(
\frac{\vec{r}_{11^{\prime}}^{\;2}\,\vec{r}_{22^{\prime}}^{\;2}-2\vec{r}%
_{12}^{\;2}\,\vec{r}_{1^{\prime}2^{\prime}}^{\;2}}{d}\ln\left(  \frac{\vec
{r}_{12^{\prime}}^{\;2}\,\vec{r}_{21^{\prime}}^{\;2}}{\vec{r}_{11^{\prime}%
}^{\;2}\vec{r}_{22^{\prime}}^{\;2}}\right)  -1\right)  +\frac{\vec{r}%
_{12}^{\;2}}{\vec{r}\,\,_{11^{\prime}}^{2}\vec{r}\,\,_{22^{\prime}}^{2}\vec
{r}_{1^{\prime}2^{\prime}}^{\;2}}\ln\left(  \frac{\vec{r}_{12}^{\;2}\vec
{r}_{1^{\prime}2^{\prime}}^{\;2}}{\vec{r}_{12^{\prime}}^{\;2}\,\vec
{r}_{21^{\prime}}^{\;2}}\right) \nonumber
\end{equation}%
\begin{equation}
+\frac{1}{2\vec{r}\,\,_{11^{\prime}}^{2}\ \vec{r}\,\,_{22^{\prime}}^{2}}%
\ln\left(  \frac{\vec{r}_{12^{\prime}}^{\;2}\,\vec{r}_{21^{\prime}}^{\;2}%
}{\vec{r}_{11^{\prime}}^{\;2}\vec{r}_{22^{\prime}}^{\;2}}\right)  \left(
\frac{\vec{r}\,\,_{12}^{4}}{d}-\frac{\vec{r}\,\,_{12}^{2}}{\vec{r}%
\,\,_{1^{\prime}2^{\prime}}^{2}}\right)  ,\quad d=\vec{r}_{12^{\prime}}%
^{\;2}\vec{r}_{21^{\prime}}^{\;2}-\vec{r}_{11^{\prime}}^{\;2}\vec
{r}_{22^{\prime}}^{\,\,2}. \label{gluon_g4}%
\end{equation}
>From the representation (\ref{K_in_corrdinate_repr}) and the expressions
(\ref{gluon_g0})--(\ref{gluon_g4}) it can be seen that the conformal
invariance is violated only by the terms proportional to $11/6$. Remind that
in the quark contribution the violation has the same form with $-n_{f}/3$
($n_{f}$ is the quark flavour number) instead of $11/6$, so that the total
violation is proportional to the $\beta$-function. It means that the kernel
$\hat{\mathcal{K}}^{QC}=\hat{\emph{K}}-[\hat{\mathcal{K}}^{B}, O_{1}]$ is
quasi-conformal, i.e. nonconformal terms in its M\"{o}bius form have origin
from the renormalization procedure.

The part of the gluon contribution to the M\"{o}bius form of $\hat
{\mathcal{K}}^{QC}$ symmetric with respect to the substitution ($\vec{r}%
_{1}^{\;\prime}\leftrightarrow\vec{r}_{2}^{\;\prime}$) coincides with the
corresponding contribution to the kernel for the evolution of the
\textquotedblleft composite dipole operators" obtained in Eq. (70) of
Ref.~\cite{Balitsky:2009xg}, if one does not pay attention to the misprint in
this equation (instead of ${d^{2}z_{3}d^{2}z_{4}}/{z_{34}^{2}}$ must be
${d^{2}z_{3}d^{2}z_{4}}/{z_{34}^{4}}$) and to the difference in the
renormalization scales: $r_{\mu_{BC}}^{2}$ instead of our $r_{\mu}^{2}$
(\ref{rmu}), being
\begin{equation}
r_{\mu_{BC}}^{2}=\frac{r_{\mu}^{2}}{4e^{2\psi(1)}},~~~~~~~\mu_{BC}^{2}%
=\frac{\mu^{2}}{4e^{2\psi(1)}}.\label{mu}%
\end{equation}
As was pointed out in Ref.~\cite{Fadin:2009za}, we think that this difference
arose because the renormalization scheme used in Refs.~\cite{Balitsky:2008zz}
and \cite{Balitsky:2009xg} is not equivalent to the conventional
$\overline{MS}$ renormalization scheme defined in the momentum space.

\section{M\"{o}bius forms for total (quasi-)conformal kernels}

In this section we present the M\"{o}bius form for the total quasi-conformal
BFKL kernels in QCD and extended supersymmetric Yang-Mills theories with
arbitrary $N$. In all these theories the quasi-conformal kernel $\hat
{\mathcal{K}}^{QC}$ is defined by the relation
\begin{equation}
\hat{\mathcal{K}}^{QC}=\hat{\mathcal{K}}+\frac{\alpha_{s}}{8\pi}\beta_{0}%
[\hat{\mathcal{K}}^{B},\ln\left(  \hat{\vec{q}}_{1}^{\;2}\hat{\vec{q}}%
_{2}^{\;2}\right)  ]-[{\hat{\mathcal{K}}\,}^{B},\hat{O}+\hat{O}_{1}%
],\label{QC kernel}%
\end{equation}
where $\beta_{0}$ is the first coefficient of the beta-function for the
corresponding theory, the operators $\hat{O}$ and $\hat{O}_{1}$ are defined in
Eqs.~(\ref{O}) and (\ref{O1}) respectively. For QCD the kernel $\mathcal{K}$
is the usual BFKL kernel defined in the momentum representation (see
Eq.~(\ref{K_momentum_repr_definition})). For SUSY Yang-Mills theories in the
$\overline{MS}$ renormalization scheme it is obtained \cite{Kotikov:2000pm}
from the QCD kernel by the change of the coefficients $n_{f}$ with $n_{M}%
N_{c}$ ($n_{M}$ is the number of gluinos, $n_{M}=N$) in the \textquotedblleft
non-Abelian" part and $n_{f}$ with $-n_{M}N_{c}^{3}$ in the \textquotedblleft
Abelian" part of the quark contribution, and by addition of the contribution
of $n_{s}$ scalars ($n_{S}=2(N-1)$). The latter contribution is defined in the
momentum space by Eqs.~(19)-(22) and (28) in Ref.~\cite{Fadin:2007xy}.

In the QCD case, using the gluon contribution from the previous section and
taking the quark part from Refs.~\cite{Fadin:2006ha,Fadin:2007ee}, for the
M\"{o}bius form of the quasi-conformal kernel (\ref{QC kernel}) we get
\begin{equation}
g^{QC(QCD)}_{0}(\vec{r}_{1},\vec{r}_{2};\vec{\rho})=6\pi\zeta\left(  3\right)
\delta\left(  \vec{\rho}\right)  -g^{QC(QCD)}_{1}(\vec{r}_{1},\vec{r}_{2}%
;\vec{\rho})~, \label{g-three-point}%
\end{equation}%
\[
g^{QC(QCD)}_{1}(\vec{r}_{1},\vec{r}_{2};\vec{r}_{2}^{\;\prime})\
\]
\begin{equation}
=\frac{\vec{r}_{12}^{\;2}}{\vec{r}_{22^{\prime}}^{\;2}\vec{r}_{12^{\prime}%
}^{\;2}}\left[  \frac{67}{18}-\zeta(2)-\frac{5n_{f}}{9N_{c}}+\frac{\beta_{0}%
}{2N_{c}}\ln\left(  \frac{\vec{r}_{12}^{\;2}\mu^{2}}{4e^{2\psi(1)}}\right)
+\frac{\beta_{0}}{2N_{c}}\frac{\vec{r}_{12^{\prime}}^{\,\,\,2}-\vec
{r}_{22^{\prime}}^{\,\,\,2}}{\vec{r}_{12}^{\;2}}\ln\left(  \frac{\vec
{r}_{22^{\prime}}^{\,\,\,2}}{\vec{r}_{12^{\prime}}^{\,\,\,2}}\right)  \right]
,
\end{equation}%
\[
g^{QC(QCD)}_{2}(\vec{r}_{1},\vec{r}_{2};\vec{r}_{1}^{\;\prime},\vec{r}%
_{2}^{\;\prime})=\frac{1}{\vec{r}_{1^{\prime}2^{\prime}}^{\,\,4}}\left(
\frac{\vec{r}_{11^{\prime}}^{\;2}\,\vec{r}_{22^{\prime}}^{\;2}-2\vec{r}%
_{12}^{\;2}\,\vec{r}_{1^{\prime}2^{\prime}}^{\;2}}{d}\ln\left(  \frac{\vec
{r}_{12^{\prime}}^{\;2}\,\vec{r}_{21^{\prime}}^{\;2}}{\vec{r}_{11^{\prime}%
}^{\;2}\vec{r}_{22^{\prime}}^{\;2}}\right)  -1\right)  \left(  1+\frac{n_{f}%
}{N_{c}^{3}}\right)
\]%
\begin{equation}
+\left(  \frac{3n_{f}}{2N_{c}^{3}}\frac{\vec{r}_{12}^{\;2}\,}{\vec
{r}_{1^{\prime}2^{\prime}}^{\,\,2}d}+\frac{1}{2\vec{r}\,\,_{11^{\prime}}%
^{2}\ \vec{r}\,\,_{22^{\prime}}^{2}}\left(  \frac{\vec{r}\,\,_{12}^{4}}%
{d}-\frac{\vec{r}\,\,_{12}^{2}}{\vec{r}\,\,_{1^{\prime}2^{\prime}}^{2}%
}\right)  \right)  \ln\left(  \frac{\vec{r}_{12^{\prime}}^{\;2}\,\vec
{r}_{21^{\prime}}^{\;2}}{\vec{r}_{11^{\prime}}^{\;2}\vec{r}_{22^{\prime}%
}^{\;2}}\right)  +\frac{\vec{r}_{12}^{\;2}}{\vec{r}\,\,_{11^{\prime}}^{2}%
\vec{r}\,\,_{22^{\prime}}^{2}\vec{r}_{1^{\prime}2^{\prime}}^{\;2}}\ln\left(
\frac{\vec{r}_{12}^{\;2}\vec{r}_{1^{\prime}2^{\prime}}^{\;2}}{\vec
{r}_{12^{\prime}}^{\;2}\,\vec{r}_{21^{\prime}}^{\;2}}\right)  .
\label{g-four-point}%
\end{equation}
Here $\beta_{0}=\frac{11}{3}N_{c}-\frac{2}{3}n_{f}$. With account of
$r_{1}\leftrightarrow r_{2}$ symmetrization this result coincides with the
result of Ref.~\cite{Balitsky:2009xg} up to the different $\mu$ (\ref{mu}).

Since the dimensional regularization violates the supersymmetry, the
regularization which is commonly used in supersymmetric theories is the
dimensional reduction. So, we take the scalar and fermion contributions from
Ref.~\cite{Fadin:2007xy}, add the gluon one found in the previous section, and
express our result in the dimensional reduction scheme, which differs from the
$\overline{MS}$ scheme by the finite charge renormalization (see
Ref.~\cite{Fadin:2007xy} for details):
\begin{equation}
\alpha_{s}\rightarrow\alpha_{s}\left(  1-\frac{\alpha_{s}N_{c}}{12\pi}\right)
. \label{dim_reduction}%
\end{equation}
Thus we find%
\begin{equation}
g_{SUSY}^{0}(\vec{r}_{1},\vec{r}_{2};\vec{\rho})=6\pi\zeta\left(  3\right)
\delta\left(  \vec{\rho}\right)  -g_{SUSY}(\vec{r}_{1},\vec{r}_{2};\vec{\rho
})~,
\end{equation}%
\[
g_{SUSY}(\vec{r}_{1},\vec{r}_{2};\vec{r}_{2}^{\;\prime})\
\]
\begin{equation}
=\frac{\vec{r}_{12}^{\;2}}{\vec{r}_{22^{\prime}}^{\;2}\vec{r}_{12^{\prime}%
}^{\;2}}\left[  \frac{32}{9}-\zeta(2)-\frac{5n_{M}+2n_{S}}{9}+\frac{\beta_{0}%
}{2N_{c}}\ln\left(  \frac{\vec{r}_{12}^{\;2}\mu^{2}}{4e^{2\psi(1)}}\right)
+\frac{\beta_{0}}{2N_{c}}\frac{\vec{r}_{12^{\prime}}^{\,\,\,2}-\vec
{r}_{22^{\prime}}^{\,\,\,2}}{\vec{r}_{12}^{\;2}}\ln\left(  \frac{\vec
{r}_{22^{\prime}}^{\,\,\,2}}{\vec{r}_{12^{\prime}}^{\,\,\,2}}\right)  \right]
,
\end{equation}%
\[
g_{SUSY}(\vec{r}_{1},\vec{r}_{2};\vec{r}_{1}^{\;\prime},\vec{r}_{2}^{\;\prime
})=\frac{1}{\vec{r}_{1^{\prime}2^{\prime}}^{\,\,4}}\left(  \frac{\vec
{r}_{11^{\prime}}^{\;2}\,\vec{r}_{22^{\prime}}^{\;2}-2\vec{r}_{12}^{\;2}%
\,\vec{r}_{1^{\prime}2^{\prime}}^{\;2}}{d}\ln\left(  \frac{\vec{r}%
_{12^{\prime}}^{\;2}\,\vec{r}_{21^{\prime}}^{\;2}}{\vec{r}_{11^{\prime}}%
^{\;2}\vec{r}_{22^{\prime}}^{\;2}}\right)  -1\right)  \left(  1-n_{M}%
+\frac{n_{S}}{2}\right)
\]%
\[
+\left(  \frac{(2n_{S}-3n_{M})}{2\vec{r}_{1^{\prime}2^{\prime}}^{\,\,2}}%
\frac{\vec{r}_{12}^{\;2}\,}{d}+\frac{1}{2\vec{r}\,\,_{11^{\prime}}^{2}%
\ \vec{r}\,\,_{22^{\prime}}^{2}}\left(  \frac{\vec{r}\,\,_{12}^{4}}{d}%
-\frac{\vec{r}\,\,_{12}^{2}}{\vec{r}\,\,_{1^{\prime}2^{\prime}}^{2}}\right)
\right)  \ln\left(  \frac{\vec{r}_{12^{\prime}}^{\;2}\,\vec{r}_{21^{\prime}%
}^{\;2}}{\vec{r}_{11^{\prime}}^{\;2}\vec{r}_{22^{\prime}}^{\;2}}\right)
+\frac{\vec{r}_{12}^{\;2}}{\vec{r}\,\,_{11^{\prime}}^{2}\vec{r}%
\,\,_{22^{\prime}}^{2}\vec{r}_{1^{\prime}2^{\prime}}^{\;2}}\ln\left(
\frac{\vec{r}_{12}^{\;2}\vec{r}_{1^{\prime}2^{\prime}}^{\;2}}{\vec
{r}_{12^{\prime}}^{\;2}\,\vec{r}_{21^{\prime}}^{\;2}}\right)  ,
\]%
\begin{equation}
\;\;d=\vec{r}_{12^{\prime}}^{\;2}\vec{r}_{21^{\prime}}^{\;2}-\vec
{r}_{11^{\prime}}^{\;2}\vec{r}_{22^{\prime}}^{\,\,2},\quad\beta_{0}=\left(
\frac{11}{3}-\frac{2n_{M}}{3}-\frac{n_{S}}{6}\right)  N_{c}~.
\end{equation}
Finally for $N=4$ SUSY theory, we put $n_{S}=6,\,\,n_{M}=4,\,\beta_{0}=0$ and
write%
\[
\langle\vec{r}_{1}\vec{r}_{2}|\hat{\mathcal{K}}^{QC}_{M}|\vec{r}_{1}%
^{\;\prime}\vec{r}_{2}^{\;\prime}\rangle{\large _{N=4} }%
\]
\[
=\frac{\alpha_{s}N_{c}}{2\pi^{2}}\int d\vec{\rho}\frac{\vec{r}_{12}{}^{2}%
}{\vec{r}_{1\rho}^{\,\,2}\vec{r}_{2\rho}^{\,\,2}}\Biggl[\delta(\vec
{r}_{11^{\prime}})\delta(\vec{r}_{2^{\prime}\rho})+\delta(\vec{r}_{1^{\prime
}\rho})\delta(\vec{r}_{22^{\prime}})-\delta(\vec{r}_{11^{\prime}})\delta
({r}_{22^{\prime}})\Biggr]~\left(  1-\frac{\alpha_{s}N_{c}\zeta(2)}{2\pi
}\right)
\]%
\begin{equation}
+\frac{\alpha_{s}^{2}N_{c}^{2}}{4\pi^{4}}\left[  \frac{\ln\left(  \frac
{\vec{r}_{12^{\prime}}^{\;2}\,\vec{r}_{21^{\prime}}^{\;2}}{\vec{r}%
_{11^{\prime}}^{\;2}\vec{r}_{22^{\prime}}^{\;2}}\right)  }{2\vec
{r}\,\,_{11^{\prime}}^{2}\ \vec{r}\,\,_{22^{\prime}}^{2}}\left(  \frac{\vec
{r}\,\,_{12}^{4}}{\vec{r}_{12^{\prime}}^{\;2}\vec{r}_{21^{\prime}}^{\;2}%
-\vec{r}_{11^{\prime}}^{\;2}\vec{r}_{22^{\prime}}^{\,\,2}}-\frac{\vec
{r}\,\,_{12}^{2}}{\vec{r}\,\,_{1^{\prime}2^{\prime}}^{2}}\right)  +\frac
{\vec{r}_{12}^{\;2}\ln\left(  \frac{\vec{r}_{12}^{\;2}\vec{r}_{1^{\prime
}2^{\prime}}^{\;2}}{\vec{r}_{12^{\prime}}^{\;2}\,\vec{r}_{21^{\prime}}^{\;2}%
}\right)  }{\vec{r}\,\,_{11^{\prime}}^{2}\vec{r}\,\,_{22^{\prime}}^{2}\vec
{r}_{1^{\prime}2^{\prime}}^{\;2}}+6\pi^{2}\zeta\left(  3\right)  \delta
(\vec{r}_{11^{\prime}})\delta({r}_{22^{\prime}})\right]  .
\end{equation}
This kernel is conformally invariant and coincides with the linearized BK
kernel obtained in Ref.~\cite{Balitsky:2009xg} with account of $r_{1}%
\leftrightarrow r_{2}$ symmetrization.

\section{Conclusion}

The main results of this paper are the following. First, we demonstrated that
the discrepancy between the gluon contribution to the M\"{o}bius form of the
BFKL kernel, calculated in Ref.~\cite{Fadin:2007de}, and the corresponding
contribution to the kernel of the colour dipole model, calculated in
Refs.~\cite{Balitsky:2008zz} and \cite{Balitsky:2009xg}, can be removed due to
the ambiguity of the kernels in the next-to-leading order, which allows the
transformations (\ref{kernel_transforms}). It was explicitly shown that the
symmetrized gluon part of the M\"{o}bius form of the kernel
(\ref{matching kernel}) coincides (up to the difference in the renormalization
scales (\ref{mu})) with the gluon part of the kernel of the colour dipole
approach found in Ref.~\cite{Balitsky:2008zz} (with account of the correction
of Ref.~\cite{Balitsky:2009xg}). In our opinion, the scales differ because the
renormalization scheme used in Refs.~\cite{Balitsky:2008zz} and
\cite{Balitsky:2009xg} is not equivalent to the conventional $\overline{MS}$
renormalization scheme defined in the momentum space.

Second, using the ambiguity mentioned above and the results of
Ref.~\cite{Balitsky:2009xg}, we constructed the quasi-conformal kernel
(\ref{QC kernel}) and found the M\"{o}bius form of this kernel in QCD and
$N$--extended supersymmetric Yang-Mills theories. The nonconformal terms in
this form are proportional to the first coefficient of the $\beta$--function.
At $N=4$ the M\"{o}bius form is conformally invariant and coincides with the
result of Ref.~\cite{Balitsky:2009xg}.

\vspace{1.0cm} \noindent{\Large \textbf{Acknowledgment}} \vspace{0.5cm}

V.S.F. thanks the Dipartimento di Fisica dell'Universit\`a della Calabria and
the Istituto Nazionale di Fisica Nucleare, Gruppo Collegato di Cosenza, for
the warm hospitality while part of this work was done and for the financial support.

\end{document}